\documentclass[MSNbibl,nameyear,dvips]{arxstspdf}
\usepackage{flushend}
\usepackage{stfloats}

\volume{26}
\issue{2}
\pubyear{2011}
\firstpage{227}
\lastpage{230}
\doi{10.1214/11-STS331A}
\referstodoi{10.1214/10-STS331}

\begin{document}
\begin{frontmatter}

\title{Discussion of ``Bayesian Models and Methods in
Public Policy and Government Settings'' by S. E. Fienberg}
\runtitle{Discussion}
\pdftitle{Discussion of Bayesian Models and Methods in
Public Policy and Government Settings by S. E. Fienberg}

\begin{aug}
\author{\fnms{David J.} \snm{Hand}\corref{}\ead[label=e1]{d.j.hand@imperial.ac.uk}%
\ead[label=u1,url]{http://www3.imperial.ac.uk/people/d.j.hand}}
\runauthor{D. J. Hand}

\affiliation{Imperial College, London}

\address{David J. Hand is Professor of Statistics, Department of Mathematics,
Imperial College London, South Kensington Campus, London SW7 2AZ, United Kingdom
\printead{e1,u1}.}

\end{aug}

\begin{abstract}
Fienberg convincingly demonstrates that Bayesian
models and methods represent a powerful approach to squeezing
illumination from data in public policy settings. However, no school of
inference is without its weaknesses, and, in the face of the
ambiguities, uncertainties, and poorly posed questions of the real
world, perhaps we should not expect to find a formally correct
inferential strategy which can be universally applied, whatever the
nature of the question: we should not expect to be able to identify a
``norm'' approach. An analogy is made between George Box's ``no models are
right, but some are useful,'' and inferential systems.
\end{abstract}


\end{frontmatter}

 Professor Fienberg has compiled an impressive collection of
examples illustrating the power of the\break Bayesian approach in public
policy and government applications. However, while these are compelling
illustrations of this power, I am uneasy about using them as the basis
for an assertion that this single approach should be adopted as the
``norm'' (which I take to mean ``should be adopted as the standard
practice''). Does it not mean, instead, that Bayesian approaches are
valuable tools to be included in the armory of every statistician
working in public policy and government settings, so that the
statistician is better able to pick an approach, method, or class of
methods which will shed most light on the problem he or she is
tackling? That is, rather than arguing for a ``norm'' method of
inference, should we not accept that it is unrealistic to hope to find
such a~single norm, and instead accept that different approaches are
suited to different situations and questions? To follow the comment in
Bayarri and Berger (\citeyear{BayBer04}), should we not recognize that ``statisticians
should readily use both Bayesian and frequentist ideas'' and ``that
each approach has a great deal to contribute to statistical practice''?
In fact, going even further than this, perhaps it is brave to assert
that a unique formal system, with clear and precise definitions and
methods, is adequate to provide an inferential mapping from the real
world, with all its uncertainties, ambiguities, inadequate definitions,
and poorly posed problems. Fienberg himself has elsewhere remarked that
``the bottom line for me involves drawing upon a~mix of pragmatism and
principle'' (Fienberg, \citeyear{Fie06}). We recognize that our statistical models
are only models, and that none are ``right'' (as George Box told us),
so why should we believe that any particular inferential strategy is
``right,'' in the sense that it should be adopted as the norm?

This point is perhaps reinforced by Fienberg's introductory critique of
the Bayesian perspective, in which he notes that the most common
criticism ``is that, since there is no single correct prior
distribution, $g( \theta)$, all conclusions drawn from the posterior
distribution are suspect.'' He says that responses to this criticism
include the recommendation that one should ``consider and report the
results associated with a variety of prior distributions'' and ``that
one should choose as a prior distribution one that in some sense
eliminates personal subjectivity.'' He does not argue that these remove
the difficulty, but he does go on to say that ``one characteristic of
the Bayesian argument that weakens this criticism$\ldots$ is that the more
data we collect, the less influence the prior distribution has on the
posterior relative to that of the data.'' This justification seems to
be equivalent to saying that, while there may be a fundamental
theoretical difficulty with Bayesian inference, this may not matter in
practice. That would mean that while it was fine as a practical
inferential tool, it also could not be regarded as a formally
``correct'' approach to inference---it would be (and indeed is) just an
approximation to the complexities of building an adequate model for
inference in the real world. After all, contrary to what is
occasionally glibly asserted, Bayesian methods do not provide a
solution to the problem of inductive inference: a Bayesian analysis
does not answer the question ``what should I believe?'' All Bayesian
methods do is provide a rational (coherent, consistent) way to update
beliefs, to change one's beliefs in the light of evidence. This is
implicit in Fienberg's comment that ``Bayesian methodology$\ldots$ provides
an internally consistent and coherent normative methodology.'' But the
internal consistency says nothing about inference beyond the formal
system involved; to do this we do need to include the prior.

In contrast to the consistency and coherence of Bayesian methodology,
Fienberg says ``frequentist methodology has no such consistent
normative fra\-mework.'' Perhaps, but on the other hand, as Cox
[(\citeyear{Cox06}),
page 197] succinctly puts it: ``Frequentist analyses are based on a
simple and powerful unifying principle. The implications of data are
examined using measuring techniques such as confidence limits and
significance tests calibrated, as are other measuring instruments,
indirectly by the hypothetical consequences of their repeated use. In
particular, they use the notion that consistency in a certain respect
of the data with a~specified value of the parameter of interest can be
assessed via a~$p$-value.'' That is, we draw conclusions using a~system
which, provided it is used properly, and subject to any assumptions it
makes, will be right most of the time---and ``being right most of the
time'' is surely a~good justification for using this approach.

Fienberg's Section 2, in which the criticisms of Bayesian methods
quoted above are given, is entitled ``The arguments for and against the
use of Bayesian methods.'' He manages to resist rehearsing the many
criticisms of frequentist approaches, possibly because they would have
made a paper in their own right, but presumably they might also have
been marshalled as arguments \textit{for} Bayesian methods (on the
principle that ``my enemy's enemy is my friend''). But he could have
listed many other criticisms of the Bayesian school, such as the Dutch
book problem, the problem of learning over time (coherence arguments
refer to a static situation), the interpretation of a probability $1/2$
as meaning the same thing whether based on a million coin tosses or
subjective opinion, multiparameter issues, as well as others. These and
other matters have been long fought over, and while many will doubtless
have been resolved to the satisfaction of at least some people, their
mere presence suggests that there are more questions about the Bayesian
strategy than is sometimes recognized. That is, it suggests that this
is merely an approach to \textit{approximating} the complexity of
inference for the natural world.

Fienberg draws attention to the incorrect notion that frequentist
methods are ``objective,'' and also notes that pragmatic Bayesian
methods ``have many subjective elements.'' Inference is an attempt to
draw some conclusion about the real world, and I agree that it would be
naive to suppose that that can be done without subjective aspects.
Since statistical inference has the prerequisite of a mapping from the
world to the formal system within which the inference is to be
conducted, it is difficult to see how that could be done without some
subjective element or arbitrary choices. I would argue that the most
important aspects of any statistical analysis are deciding what the
scientific question is, and then making an effective (or at least
adequate) mapping to a statistical question (e.g., Hand, \citeyear{Han94}, \citeyear{HAN96}).
Moreover, since such mappings always lead to formal representations
which are, at best, only approximations to the (perhaps even typically
rather ill-posed) scientific question, too much emphasis on the
niceties of the formal statistical inferential method may be
unnecessary. In my view, far too little attention is devoted, in both
statistical education and practice, to the key issue of establishing
the mapping---before the formal inferential tools can be applied.

Perhaps the most impressive thing about Fienberg's choice of examples
is their scope, both in terms of application areas and in terms of the
way\vadjust{\eject} in which Bayesian methodology is applied. I will make a few
comments on the examples below, but it would be naive to expect me to
be able to fault the analyses, on two counts. First, the examples were
presumably chosen as exemplars of the effectiveness of such methodology
in such applications, and this is obviously best done by choosing
examples where the methodology is indeed effective. And second, the
sensitive and sophisticated use of any statistical school (in this case
a particular one) by someone who really knows what he or she is doing
is likely to lead to effective results. By ``know what he or she is
doing,'' I mean balancing pragmatism and principles, as noted above.
This includes the ability to choose a model form which captures the
essence of the problem, rather than either simply trying to model every
aspect of the underlying system or failing to include some vital
aspect. The first of these failures would risk rejecting a model which
could be perfectly adequate for the purpose at hand, even though it
failed to match the data in other ways, and the second would mean the
model was not fit for the purpose. This is all part of the art of
statistics.

I think, therefore, that the impressiveness of the examples does not
establish that this approach\break should be adopted as the ``norm,'' but
rather that such an approach is highly effective (for these kinds of
problems), when used by someone who knows what he or she is doing.

Several of the examples (small area estimation and census adjustment,
election night forecasting, post-marketing surveillance of drugs) hinge
on the notion of ``borrowing strength.'' This is a powerful tool, built
on so-called empirical Bayes, which really represents something of a
Bayesian frequentist synthesis. I am sure we will see such methods used
more and more often as electronic data capture facilitates the easy
compilation of massive data sets, permitting exploration of finer and
finer partitions. Continuing debates about national censuses (e.g.,
Canada's abandonment of the long-form census) mean that such
sophisticated tools are likely to have an important public policy role
in the future. Likewise, the growing use of league tables and other
systems for ranking and rating hospitals (even individual surgeons),
schools, local authorities, police forces, and other public bodies are
areas where such adjustments can be very valuable. At a more refined
level, the problem of detecting adverse drug reactions in
post-marketing surveillance of drugs is an example of a type of problem
which produces\vadjust{\eject} a very large number of cells arranged in a
cross-classification---and, naturally, often relatively small cell
counts. Overcoming this by borrowing strength requires some way of
determining the ``similarity'' between cells. This might be done in
various ways---assuming independence between the factors of the
cross-classification, using external information characterizing the
objects, or based on measures of similarities between the row and
column count profiles (e.g., Zhang, \citeyear{ZHA}).

Other examples, however, cover quite different\break kinds of application of
Bayesian methods---and so reveal the strengths of Bayesian approaches
in quite different ways. One such area which is relatively new is that
of adaptive clinical trials. If borrowing strength has a natural
Bayesian interpretation, so too does adaptive allocation of patients,
as one's state of knowledge changes.

Statistics might be defined as the science of uncertainty, so
Fienberg's example of climate change is very fitting---and perhaps this
is an example which does very naturally fall into a Bayesian mold. The
UK's Royal Society has recently produced a summary of the current
scientific evidence on climate change and its drivers, which spells out
``where the science is well established, where there is wide consensus
but continuing debate, and where there remains substantial
uncertainty'' (Royal Society,\break \citeyear{autokey7}).

One of the recognized strengths of Bayesian methods is that they can
extract information from small samples (albeit at the cost of using
``information'' from other sources), but that does not mean they are
restricted to small samples; indeed the potential to produce
cross-classifications of large samples, noted above, means that the
demand for Bayesian tools can be just as great with large samples.
Fienberg does not explicitly use the phrase ``data mining'' (though
it appears twice in his references). However, his final paragraph draws
attention to the use of latent variable models with very large data
sets. More generally, data mining has made extensive use of hidden
Markov models, and has adopted various other Bayesian modeling
approaches, such as the use of graphical models (``Bayesian belief
networks'') in health surveillance. Having said that, I think it is
true that most work which is explicitly labeled as data mining still
has a tendency to be focused on algorithms rather than inference. The
large sample inferential work seems to be being carried out by
statisticians (e.g., Efron, \citeyear{Efr10}).

In summary, I think Fienberg goes too far in suggesting that Bayesian
methods should become the ``norm'' in public settings. Rather, I think
we should accept that no inferential system will always be appropriate,
therefore being adopted as a norm or standard approach. Instead we
should acknowledge that different systems are different strategies for
tackling a problem which defies a ``correct'' solution; that we should
therefore retain our flexibility, and match our system to our
objective, just as different models are suited to answering different
questions. However, the examples Fienberg has chosen certainly
illustrate the power of the Bayesian perspective in public policy
applications. The examples also illustrate how advances in statistics
are driven by practical applications, and the historical backgrounds to
the examples also show rather beautifully how statistical ideas develop
over time. I thoroughly enjoyed the paper.
\vspace*{-6pt}


\end{document}